\documentclass[a4paper,fleqn,usenatbib,useAMS]{mnras}

\usepackage{graphicx}	
\usepackage{amsmath}	
\usepackage{amssymb}	
\usepackage{multicol}        
\usepackage{bm}		
\usepackage{pdflscape}	
\usepackage{gensymb}
\usepackage[normalem]{ulem}

\renewcommand{\eqref}[1]{Eq.~(\ref{#1})}

\newcommand{\bea}{\begin{eqnarray}}
\newcommand{\eea}{\end{eqnarray}}
\newcommand{\beq}{\begin{equation}}
\newcommand{\eeq}{\end{equation}}

\newcommand{\tA}{\tau_{\rm A}}

\newcommand{\tnl}{\tau_{\rm nl}}

\newcommand{\vskipfig}{\vskip-0.25cm}
\newcommand{\dd}{\partial}
\newcommand{\vB}{\mathbf{B}}

\newcommand{\vu}{\mathbf{u}}
\newcommand{\vb}{\mathbf{b}}

\newcommand{\vz}{\mathbf{z}}

\newcommand{\vup}{\vu_\perp}
\newcommand{\vbp}{\vb_\perp}
\newcommand{\vzp}{\vz_\perp}

\newcommand{\dzp}{\delta z_\perp}

\newcommand{\dvzp}{\delta \vz_\perp}

\newcommand{\lpar}{l_\parallel}

\newcommand{\vrpp}{\mathbf{r}_\perp}

\usepackage[T1]{fontenc}
\usepackage{ae,aecompl}

\usepackage{times}

\title[Measures of 3D Anisotropy and Intermittency in Strong Alfv\'enic Turbulence]{Measures of Three-Dimensional Anisotropy and Intermittency in Strong Alfv\'enic Turbulence}

\author[A. Mallet et al.]{A. Mallet$^{1,2}$\thanks{Contact e-mail: \href{mailto:alfred.mallet@unh.edu}{alfred.mallet@unh.edu}}, A. A. Schekochihin$^{2,3}$, B. D. G. Chandran$^{1,3}$, C. H. K. Chen$^{4}$, 
\newauthor
T. S. Horbury$^{4}$, R. T. Wicks$^{5}$ and C. C. Greenan$^{6}$
\\
$^{1}$Space Science Center, University of New Hampshire, Durham, NH 03824, USA \\
$^{2}$Rudolf Peierls Centre for Theoretical Physics, University of Oxford, Oxford OX1 3NP, United Kingdom\\
$^{3}$Merton College, Oxford OX1 4JD, United Kingdom\\
$^{4}$Department of Physics, Imperial College London, London SW7 2AZ, United Kingdom\\
$^{5}$Institute for Risk and Disaster Reduction, University College London, London, WC1E 6BT, United Kingdom\\
$^{6}$Department of Statistics, University of Oxford, Oxford OX1 3TG, United Kingdom
}

\pubyear{2015}

\begin{document}
\label{firstpage}
\pagerange{\pageref{firstpage}--\pageref{lastpage}}
\maketitle

\begin{abstract}
We measure the local anisotropy of numerically simulated strong Alfv\'enic turbulence with respect to two local, physically relevant directions: along the local mean magnetic field and along the local direction of one of the fluctuating Elsasser fields. We find significant scaling anisotropy with respect to both these directions: the fluctuations are ``ribbon-like" --- statistically, they are elongated along both the mean magnetic field and the fluctuating field. The latter form of anisotropy is due to scale-dependent alignment of the fluctuating fields.
The intermittent scalings of the $n$th-order conditional structure functions in the direction perpendicular to both the local mean field and the fluctuations agree well with the theory of \cite{Chandran14}, while the parallel scalings are consistent with those implied bythe critical-balance conjecture. 
We quantify the relationship between the perpendicular scalings and those in the fluctuation and parallel directions, and find that the scaling exponent of the perpendicular anisotropy (i.e., of the aspect ratio of the Alfv\'enic structures in the plane perpendicular to the mean magnetic field) depends on the amplitude of the fluctuations. This is shown to be equivalent to the anticorrelation of fluctuation amplitude and alignment at each scale. 
The dependence of the anisotropy on amplitude is shown to be more significant for the anisotropy between the perpendicular and fluctuation-direction scales than it is between the perpendicular and parallel scales.
\end{abstract}




\section{Introduction}
Strong plasma turbulence is present in a wide range of astrophysical systems, and is directly measured by spacecraft in the solar wind \citep[e.g.,][]{bruno2013}. In the presence of a strong mean magnetic field $\vB_0$, on scales longer than the ion gyroradius, the Alfv\'enically polarized fluctuations decouple from the compressive fluctuations and satisfy the reduced magnetohydrodynamic (RMHD) equations. These can be derived both as an anisotropic limit of standard MHD \citep{strauss1976,kadomtsev1974} and as a large-scale limit of gyrokinetics \citep{schektome2009}, meaning that they describe the turbulence in both strongly and weakly collisional plasmas. Written using \cite{elsasser} variables $\vzp^\pm = \vup \pm \vbp$, where $\vup$ and $\vbp$ are the velocity and magnetic-field 
(in velocity units) perturbations perpendicular to $\vB_0$, the RMHD equations are
\beq
\dd_t \vzp^\pm \mp v_{\rm A} \dd_z \vzp^\pm + \vzp^\mp \cdot \nabla_\perp \vzp^\pm = -\nabla_\perp p,
\label{eq:RMHD}
\eeq
where the pressure $p$ is determined from $\nabla_\perp\cdot \vz^\pm = 0$, $v_{\rm A} = |\vB_0|$ is the Alfv\'en speed, and we have taken $\vB_0$ to be in the $z$ direction.

The turbulence described by Eqs. (\ref{eq:RMHD}) is known to be anisotropic with respect to the local magnetic-field direction, in both numerical simulations and in the solar wind \citep{chovishniac,marongoldreich,horanis,podesta,wicks2010,chenmallet,beresnyakanis}, with the anisotropy increasing at smaller scales. This anisotropy is explained by the critical-balance conjecture \citep{gs95,gs97}, which posits that the nonlinear time $\tnl^\pm$ and Alfv\'en (linear) time $\tA^\pm \doteq l_\parallel^\pm/v_{\rm A}$ must be comparable at each scale, where $l_\parallel$  is the coherence length along the magnetic field lines. The dynamics of weak turbulence ($\tA^\pm \ll \tnl^\pm$)  lead to a decrease in $\tnl^\pm$ until $\tA^\pm \sim \tnl^\pm$, while if $\tA \gg \tnl$, it is causally impossible to maintain the parallel coherence over length $l_\parallel$, so $l_\parallel$ --- and thus $\tA$ --- adjust until $\tA^\pm\sim\tnl^\pm$ \citep{gs97,nazarenko2011}. This guarantees that the two timescales are comparable, and so the cascade time is, inevitably, $\tau_\mathrm{c} \sim \tA^\pm \sim \tnl^\pm$. By an argument following \cite{k41}, the scale independence of the mean energy flux,
\beq
\varepsilon^\pm \sim \frac{(\dzp^\pm)^2}{\tau_\mathrm{c}} 
\sim \frac{(\dzp^\pm)^2 v_{\rm A}}{\lpar^\pm}
\sim \text{const},
\label{eq:eps}
\eeq
implies that $(\dzp^\pm)^2 \sim \lpar^\pm (\varepsilon^\pm / v_{\rm A})$, or, equivalently, the energy spectra of the Elsasser fields have a spectral index in the parallel direction of $-2$, regardless of the details of the nonlinear term. This is seen in both measurements of the solar wind and simulations cited above.

The perpendicular scaling is harder to establish because only $\vzp^\pm$ that has a gradient in the direction of $\vzp^\mp$ gives rise to a nonzero contribution to the RMHD nonlinearity, $\vzp^\mp \cdot \nabla_\perp \vzp^\pm$. Combined with the fact that the Elsasser-fields are 2D-solenoidal, $\nabla_\perp \cdot \vzp^\pm = 0$, this means that \emph{dynamic alignment} \citep{boldyrev} of their fluctuation vectors to within a small angle $\theta^\pm$ of each other will decrease the nonlinearity by a factor $\sin\theta^\pm$. The nonlinear time may, therefore, be defined as
\beq
\tnl^\pm \doteq \frac{\lambda}{\delta z^\mp_\perp \sin\theta^\pm},\label{eq:tnl}
\eeq
where $\lambda$ is the perpendicular coherence length.
If $\theta$ is correlated with amplitude in a scale-dependent manner, this can alter the scaling behaviour of the nonlinear time, and, therefore, the scaling of the fluctuation amplitudes. There is continuing disagreement as to whether the numerical evidence that supports the scale-dependence of the dynamic alignment angle $\sin\theta^\pm$ is truly representative of the asymptotic state of the RMHD inertial range \citep{perez14,Beresnyak14}.

The alignment of the fields and the consequent reduction in the nonlinearity can also be linked to anisotropy within the perpendicular plane \citep{boldyrev}\footnote{The argument that follows only applies to sheetlike structures. Aligned circular structures are also possible \citep{perezchandran2013}, but sheets have been observed as the dominant structures in MHD turbulence in a wide range of studies \citep{grauer1994,politano1995,marongoldreich,greco2010}. Recently, \cite{howes2015} has shown that the Alfv\'en wave dynamics lead naturally to the formation of sheetlike structures, and so our restriction to this type of structures is motivated by analysis of direct numerical simulations of the turbulence.}. Critical balance implies that 
\beq
\frac{l_\parallel}{v_\mathrm{A}} \sim \frac{\lambda}{\delta z_\perp^\pm \sin\theta^\pm},\label{eq:cb_align}
\eeq
where $l_\parallel$ is now taken to be the coherence length along the magnetic field of the combination of the fluctuating fields which make up the structure, $\vzp^+$ and $\vzp^-$. 
Meanwhile, the distance that the magnetic field lines wander in the perpendicular plane is typically of order
\beq
\xi \sim \frac{\mathrm{max}(\delta z^+,\delta z^-)}{v_\mathrm{A}} l_\parallel ,\label{eq:wander}
\eeq
where we choose the maximum of the two Elsasser fields because $b_\perp \approx z_\perp^\pm / 2$ when $z_\perp^\pm \gg z_\perp^\mp$. 
Since $l_\parallel$ is the coherence length along the field line, the combined $\vzp^+$ and $\vzp^-$ fluctuations must also be coherent \emph{in their own direction} (the "fluctuation direction") over at least the distance $\xi$. This direction is defined to within an angle $\theta^\pm$, because the fields are aligned with each other within that angle. Therefore, the typical aspect ratio of coherent structures within the perpendicular plane is $\lambda /\xi$. Comparing Eqs. (\ref{eq:cb_align}) and (\ref{eq:wander}), we find that
\beq
\sin\theta^\pm \sim \frac{\lambda}{\xi} \doteq \sin\theta.\label{eq:anisalirelation}
\eeq
The same argument was used by \cite{boldyrev} for the angle between $\delta \mathbf{u}_\perp$ and $\delta \mathbf{b}_\perp$, $\theta^{ub}$, instead of $\theta^\pm$: either angle being small reduces the nonlinearity, so whichever is smaller in the sheetlike structures will generally constrain the aspect ratio $\lambda/\xi$ better.

Combined with the anisotropy in the parallel direction, the above argument implies that the turbulence may exhibit 3D anisotropy in an instantaneous local basis defined by the directions of the mean magnetic field, the fluctuations, and the direction perpendicular to both. Equivalently, turbulent fluctuations may have different coherence scales $l_\parallel$, $\xi$, and $\lambda$ in these three directions\footnote{One expects some degree of anisotropy within the perpendicular plane just due to kinematic constraints imposed by the solenoidality of the fields $\mathbf{z}^\pm_\perp$. We discuss this issue in the Appendix, showing that solenoidality does not directly constrain the conditional structure function.}.

It is not hard to show that scale-dependent perpendicular anisotropy cannot exist without non-self-similar scale dependence of the joint distribution of the vector field-increments, i.e., without intermittency. Suppose the joint distribution $p(\mathbf{\delta z^\pm_\perp}|r_\perp)$ were invariant when the amplitude were rescaled by $r_\perp^a$, i.e., the rescaled vector variable $\mathbf{w} = \mathbf{\delta z^\pm_\perp}/r_\perp^a$ had a distribution that did not depend on $r_\perp$. The fact that the whole joint distribution is invariant means that not only are the amplitudes non-intermittent, but the angle 
\beq
\theta_{\delta z^\pm_\perp} = \arctan\frac{\delta z^\pm_{\perp x}}{\delta z^\pm_{\perp y}} = \arctan\frac{w_x}{w_y}
\eeq
also has a distribution independent of $r_\perp$. This guarantees that the conditional $n$th-order structure function has an angle-independent scaling:
\beq
S_{n,\rm 3D} = \langle (\delta z^\pm_\perp)^n | \theta_{\delta z^\pm_\perp}, r_\perp \rangle=r_\perp^{na} \langle w^n |\theta_{\delta z^\pm_\perp}\rangle = r_\perp^{na} f_n(\theta_{\delta z^\pm_\perp}),
\eeq
where the unknown function $f_n$ cannot depend on $r_\perp$. Thus, if the vector $\mathbf{\delta z^\pm_\perp}$ is non-intermittent (has a scale-invariant distribution), it cannot have scale-dependent perpendicular anisotropy or equivalently, according to the argument earlier in this Introduction, scale-dependent alignment.

In this paper, we study the 3D anisotropy and intermittency in numerically simulated RMHD turbulence, using a 3D conditional structure function method, described in Section \ref{sec:3dsf}, which was first used by \cite{chen3d} for measurements in the solar wind. In Section \ref{sec:2ord}, we present the results obtained using second-order conditional structure functions, showing that there is indeed significant 3D anisotropy. In Section \ref{sec:3dinterm}, we go further, and present the results of the 3D conditional structure function analysis for structure functions of up to 5th order, showing that the turbulence is highly intermittent in all three directions, and comparing the scalings in the perpendicular direction to a recent theoretical model of intermittency in Alfv\'enic turbulence \citep{Chandran14}, finding that the measurements are consistent with this model. The scalings in the parallel and fluctuation directions are compared to a simple model where anisotropies 
do not depend on amplitude at a particular scale, which turns out to be slightly inconsistent with the data. This implies that the anisotropy 
 is itself intermittent. 
In Section \ref{sec:ass}, we present a quantitative analysis of this \emph{intermittency of anisotropy}, and show that the scaling exponents of the 
aspect ratios $\lambda/\xi$ and $\lambda/l_\parallel$ increase with the order $n$ of the structure functions that one uses to calculate them.  We show that the perpendicular 
aspect ratio $\sin\theta \doteq \lambda /\xi$ (the anisotropy within the perpendicular plane) is significantly intermittent, while the parallel aspect ratio $\sin\phi \doteq \lambda / l_\parallel$ is less so. 
We then discuss what implications this has for the physics of the collisions of balanced Alfv\'enic fluctuations in the model of \cite{Chandran14}.
In Section \ref{sec:comp}, we compare our results on the intermittency of the anisotropy within the perpendicular plane to the scaling of the alignment angle defined more traditionally in terms of the ratio of different structure functions \citep{mcatbolalign}, and conclude that they are consistent with each other, which suggests that the two methods are indeed measuring the same phenomenon. In Section \ref{sec:disc}, we summarise our conclusions and discuss the relationship between this and previous work.

\begin{figure*}
\begin{tabular}{cc}
\includegraphics[width=8.9cm]{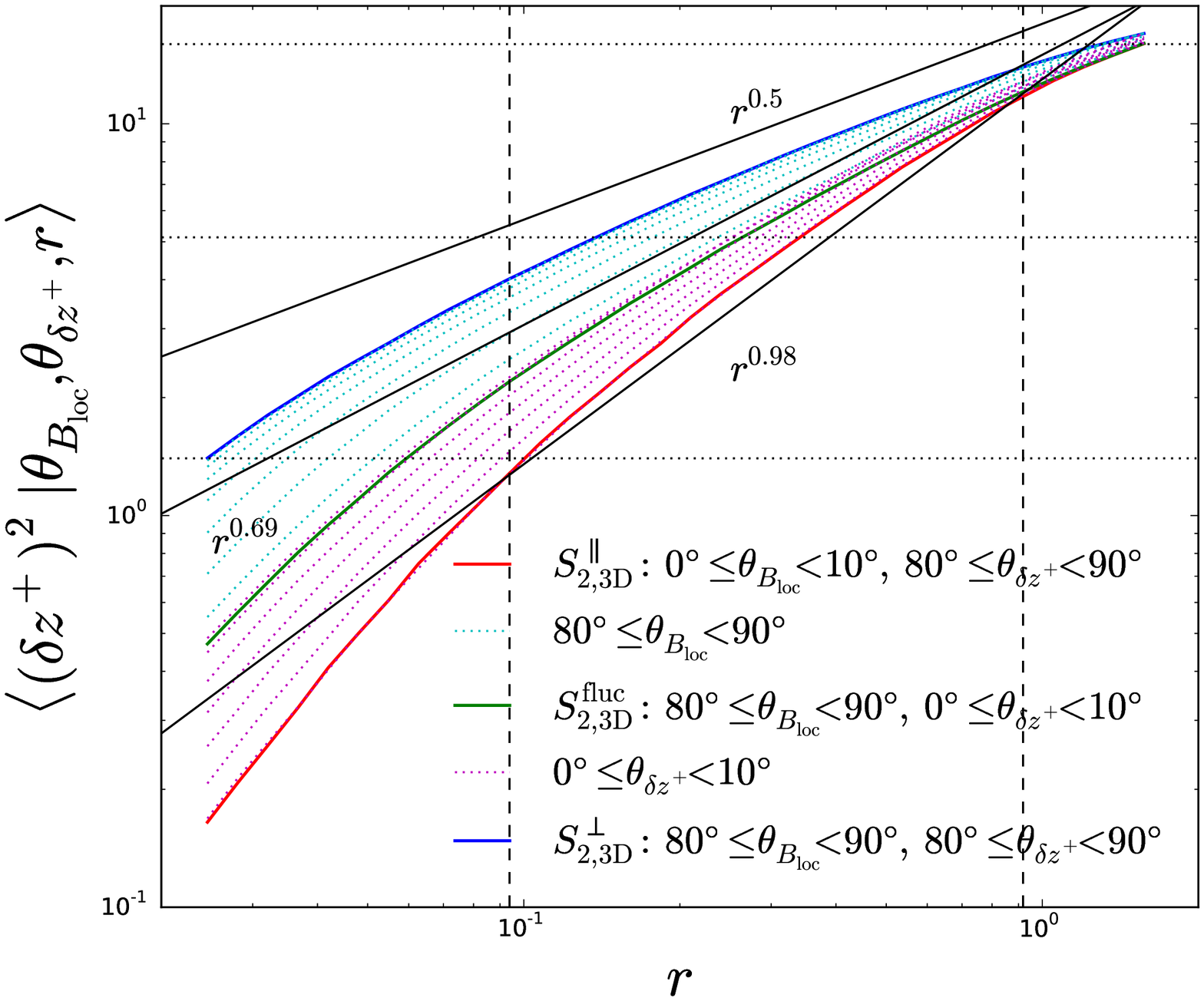} &
\includegraphics[width=8.9cm]{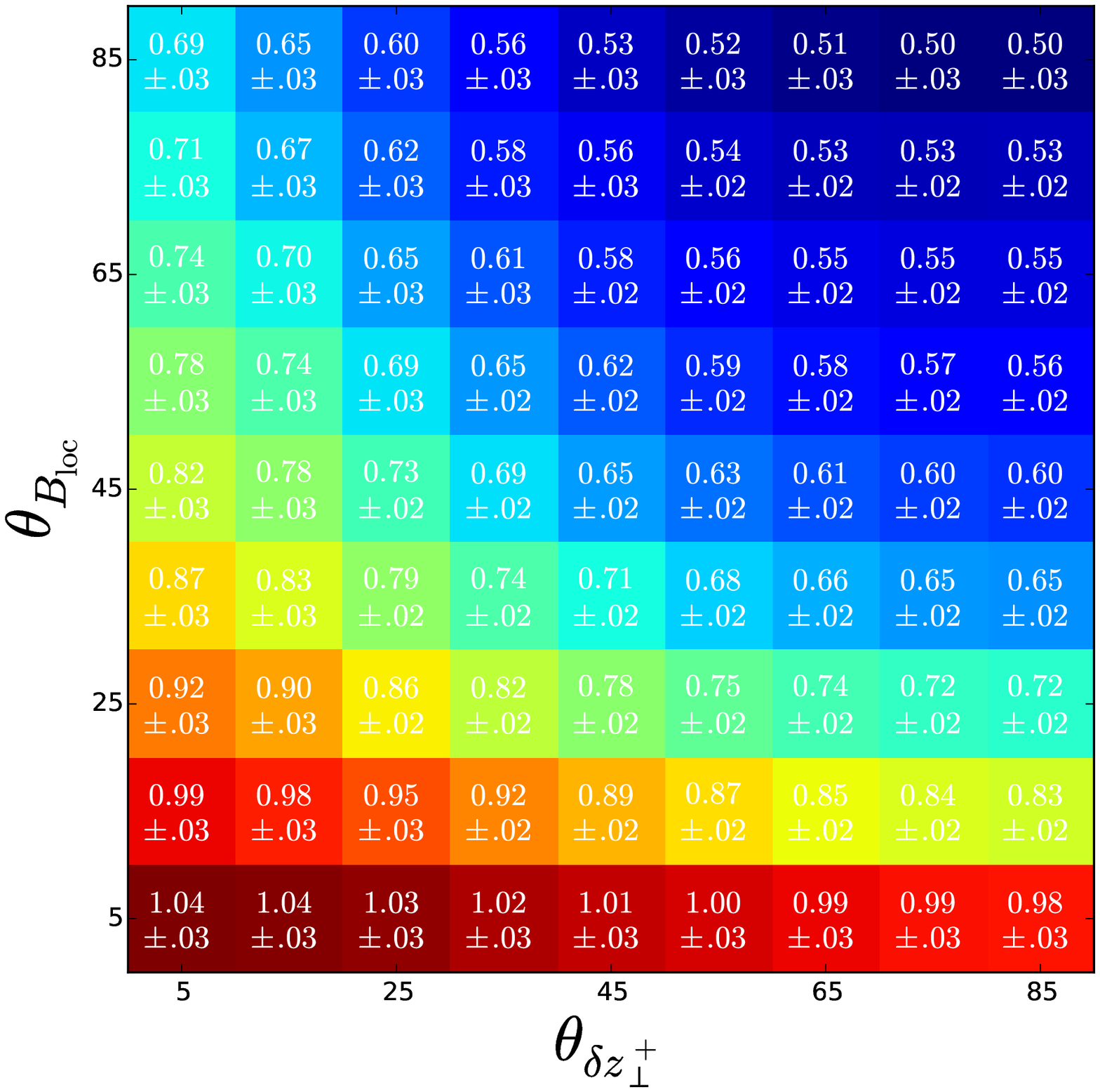} \\
\textbf{(a)} & \textbf{(b)} \\
\end{tabular}
\caption{\textbf{(a)} The conditional second-order structure functions [Eq. (\ref{eq:s3d})] plotted for different angle intervals. In dark blue (solid line) is the structure function $S_{2,\rm 3D}^\perp$ in the angle bin most perpendicular to both the local mean magnetic field and the $\delta \vzp^+$ fluctuations, in green (solid line), the structure function $S_{2,\rm 3D}^\mathrm{fluc}$ in the angle bin closest to the direction $\delta \hat{\vz}_\perp^+$, and in red (solid line), the structure function $S_{2,\rm 3D}^\parallel$ in the angle bin closest to the parallel direction $\hat{\mathbf{B}}_{\rm loc}$. In cyan (dotted lines), are the structure functions in the whole range of $\theta_{\delta \vzp^+}$ bins, with the angle to the mean field $80\degree \leq\theta_{B_{\rm loc}} < 90\degree$, and in magenta (dotted lines), the structure functions in the whole range of $\theta_{B_{\rm loc}}$ bins, but with the angle to the fluctuation $0\degree \leq \theta_{\delta z^+}\leq 10\degree$ bin. The dotted horizontal lines show the values of the structure function for which the ``statistical eddies" in Figure \ref{fig:eddy} were calculated, while the vertical dashed lines show the range over which the structure-function scaling exponents were measured. \textbf{(b)} The second-order structure function exponents $\zeta_2(\theta_{B_{\rm loc}},\theta_{\delta z^\pm_\perp})$ and the associated errors, in all angle bins. It should be noted that to compare directly with a real situation (e.g., the solar wind), the angle $\theta_{B_{\rm loc}}$ must be rescaled assuming a specific physical aspect ratio of the simulation box rather than the nominal aspect ratio of 1 [see Eq. (\ref{eq:rescaling})]. \label{fig:s2allang}}
\vskipfig
\end{figure*}
\section{Numerical setup and 3D conditional structure functions}\label{sec:3dsf}

Eqs. (\ref{eq:RMHD}) were solved in a triply periodic box of resolution $1024^3$, using the code described in \citet{chenmallet}. In the code units, $v_{\rm A} = 1$ and the box has length $2\pi$ in each direction. The RMHD equations are invariant under the simultaneous rescaling 
\beq
z\rightarrow a z,\, v_{\rm A} \rightarrow a v_{\rm A} \label{eq:rescaling}
\eeq
for arbitrary $a$. Therefore, while in code units $z^\pm_\perp \sim v_{\rm A}$ and the box is cubic, in fact, when translated into physical units, the box is much longer in the parallel direction and the fluctuation amplitudes are much smaller than $v_{\rm A}$, even as the linear and nonlinear terms remain comparable. Energy was injected via white-noise forcing at $k_\perp = 1,2$ and $k_\parallel =1$ and dissipated by perpendicular hyperviscosity ($\nu_\perp \nabla_\perp^8$ with $\nu_\perp = 2 \times 10^{-17}$). There is also an effective Laplacian parallel viscosity $\nu_\parallel=1.5\times 10^{-4}$ because the linear term is upwinded slightly; $\nu_\parallel$ is chosen to be small enough so that it only dissipates a small fraction ($\approx 7\%$) of the total power. The mean injected power was  taken to be $\varepsilon^\pm = 1$, meaning that the turbulence is balanced and strong. The forcing term is purely in the velocity, and the magnetic field was not forced so as not to break the magnetic-flux conservation at the forcing scales. 

We define the Elsasser-field increments as 
\beq
\delta \vzp^\pm = \vzp^\pm(\mathbf{r}_0 + \mathbf{r}) -\vzp^\pm(\mathbf{r}_0),\label{eq:incr}
\eeq
where $\mathbf{r}_0$ is an arbitrary point (arbitrary because we consider homogeneous turbulence), and $\mathbf{r}$ is the separation vector, with length $r$ and direction $\hat{\mathbf{r}}=\mathbf{r}/r$. The amplitude of the field increment in Eq.~(\ref{eq:incr}) is $\dzp^\pm = |\delta \vzp^\pm|$, and its direction is $\delta \hat{\vz}_\perp^\pm =  \dvzp^\pm/\dzp^\pm$. The local mean magnetic field $\mathbf{B}_{\rm loc}$ between $\mathbf{r}_0$ and $\mathbf{r}_0+\mathbf{r}$ is defined as 
\beq
\mathbf{B}_{\rm loc} = \vB_0 + \frac{1}{2}\left[\vbp(\mathbf{r}_0) + \vbp(\mathbf{r}_0+\mathbf{r})\right], \label{eq:thbloc}
\eeq
and its direction is $\hat{\vB}_{\rm loc} = \vB_{\rm loc} / |\vB_{\rm loc}|$.
The components of the field increment and the separation vector in the plane normal to $\vB_{\rm loc}$ are
\beq
\begin{split}
\delta \vz_{\perp,\rm N}^\pm &= \dvzp^\pm -  [\dvzp^\pm \cdot \hat{\vB}_{\rm loc}] \hat{\vB}_{\rm loc}, \\
\mathbf{r}_\perp &= \mathbf{r} - [\mathbf{r}\cdot\hat{\vB}_{\rm loc} ]\hat{\vB}_{\rm loc},
\end{split}
\eeq
and the directions of these vectors are $\delta \hat{\vz}_{\perp,\rm N}^\pm=\delta \vz_{\perp,\rm N}^\pm/|\delta \vz_{\perp,\rm N}^\pm|$ and $\hat{\mathbf{r}}_{\perp} = \mathbf{r}_{\perp}/|\mathbf{r}_{\perp}|$.

The angle between $\mathbf{r}$ and the local mean field is defined via
\beq
\cos\theta_{B_{\rm loc}} = \hat{\mathbf{r}}\cdot \hat{\vB}_{\rm loc}.
\eeq
It is important to point out that this angle is not invariant to the rescaling in Eq. (\ref{eq:rescaling}), and so, to compare the dependence of the structure functions (Figure \ref{fig:s2allang}(b)) on this angle to a situation with a given aspect ratio (or fluctuation level), one must rescale it assuming some specific aspect ratio $a$ [see Eq.~(\ref{eq:rescaling})] of the physical box, rather than the nominal value of 1 used in RMHD simulations. However, $\theta_{B_{\rm loc}}=0\degree,90\degree$ are fixed points under any such rescaling. The angle between $\mathbf{r}_\perp$ and the perpendicular fluctuation $\delta \vzp^\pm$ is defined via
\beq
\cos\theta_{\delta z^\pm_\perp} = \hat{\mathbf{r}}_\perp \cdot \delta \hat{\vz}_{\perp,\rm N}^\pm.
\eeq
If $\theta_{B_{\rm loc}} = 90\degree$ and $\theta_{\delta z^\pm_\perp} = 0\degree$, then the point separation $\mathbf{r}$ is along the "fluctuation direction", while if $\theta_{B_{\rm loc}} = 90\degree$ and $\theta_{\delta z_\perp^\pm} = 90\degree$, it is along the direction perpendicular to both the fluctuation and the local mean field, which we will call the "perpendicular direction". If $\theta_{B_{\rm loc}} = 0\degree$, the separation is along the "parallel direction". The angles $\theta_{B_{\rm loc}}$ and $\theta_{\delta z_\perp^\pm}$, along with the point separation $r$, define a locally-varying coordinate system referred to the two directions that we expect to be physically important\footnote{It is important to distinguish between the many angles (and aspect ratios) defined in this paper: $\theta^\pm, \theta^{ub}$ are the angles between field increments, $\sin\theta$ is the aspect ratio of structures in the local basis, and $\theta_{B_{\rm loc}}, \theta_{\delta z_\perp^\pm}$ are angles describing the relative arrangement of fields and separation vectors $\mathbf{r}$. They are not necessarily the same, although we have argued that $\theta$ is determined by the smaller of $\theta^\pm$ and $\theta^{ub}$.}.

The $n$th-order conditional structure function of $\mathbf{z}^\pm_\perp$ at point separation $r$ and the pair of angles $\theta_{B_{\rm loc}},\theta_{\delta z^\pm_\perp}$,
\beq
S_{n,\rm{3D}}(\theta_{B_{\rm loc}},\theta_{\delta z^\pm_\perp},r) = \langle (\delta z^\pm_\perp)^n | \theta_{B_{\rm loc}}, \theta_{\delta z_\perp^\pm}, r\rangle,\label{eq:s3d}
\eeq
 is defined as the average of $(\delta z^\pm_\perp)^n$ at the scale $r$, with the separation vector characterized by angles $\theta_{B_{\rm loc}}$ and $\theta_{\delta z^\pm_\perp}$. These objects (with $n=2$) have been used by \cite{chen3d} for analysis of the real solar wind turbulence.
The conditional structure function defines the scaling of the fluctuations at all angles to the physically distinct directions identified above, and provides a natural way to study the anisotropy in all directions using the same mathematical object. Our subsequent analysis is based on the calculation of these structure functions using data from the numerical simulation described above.

To achieve this, snapshots of the fields in the simulation were taken at 10 different times separated by more than a turnover time, viz., every 2 code time units. For each of the snapshots, $8 \times 10^6$ pairs of points $\mathbf{r}_0$, $\mathbf{r}_0+\mathbf{r}$ were chosen at each of 32 different logarithmically spaced separation scales $r$. The direction $\hat{\mathbf{r}}$ was uniformly distributed over a sphere. For each pair of points, the Elsasser-field increment amplitudes $\dzp^\pm$ and the three
 angles $\theta_{B_{\rm loc}}, \theta_{\delta z^\pm_\perp}$ were recorded. All angles were collapsed onto the interval $[0\degree,90\degree]$. The structure-function values reported here are the means over all 10 snapshots, and the error bars show the standard deviation from the means calculated for each snapshot. 

To calculate the $n$-th order conditional structure functions in Eq. (\ref{eq:s3d}), we bin the field-increment amplitudes $\delta z^\pm_\perp$ by the pair of angles $\theta_{B_{\rm loc}},\theta_{\delta z^\pm_\perp}$. Here we will only show the structure functions of $\delta z^+_\perp$; the $\delta z^-_\perp$ structure functions are the same because the turbulence is balanced. The conditional average in Eq. (\ref{eq:s3d}) was calculated over an angle bin $10(i-1)\degree\leq \theta_{B_{\rm loc}} <10i\degree$, $10(j-1)\degree\leq \theta_{\delta z_\perp^\pm} < 10j\degree$, where $i$ and $j$ range from 1 to 9. Some special cases of this structure function deserve particular attention and particular notation:
\beq
\begin{split}
i=1&: \quad S_{n,\rm 3D}^\parallel, \quad \text{``parallel" structure function},\nonumber\\
i=9,j=1&: \quad S_{n,\rm 3D}^\mathrm{fluc}, \quad \text{``fluctuation-direction" structure function},\nonumber\\
i=9,j=9&: \quad S_{n,\rm 3D}^\perp, \quad\text{``perpendicular" structure function}.\nonumber
\end{split}
\eeq
These bins correspond to fluctuations aligned most closely with the physical directions $\hat{\vB}_{\rm loc}$ (parallel), $\delta\hat{\mathbf{z}}^\pm_\perp$ (fluctuation), and $\hat{\vB}_{\rm loc}\times \delta\hat{\mathbf{z}}^\pm_\perp$ (perpendicular). We will refer to the scales at which those particular structure functions are sampled as the parallel scale $l_\parallel$, fluctuation-direction scale $\xi$, and perpendicular scale $\lambda$, respectively. 

\section{Second-order conditional structure functions}\label{sec:2ord}
Figure \ref{fig:s2allang} shows the conditional second-order structure functions at different angles $\theta_{B_{\rm loc}}$ and $\theta_{\delta z^+}$. These structure functions were fit to power laws over the inertial range, defined here as $0.09<r<0.92$. We define $\zeta_n(\theta_{B_{\rm loc}},\theta_{\delta z^\pm_\perp})$ as the scaling exponent of the $n$th-order structure function at that pair of angles, viz.,
\beq
S_{n,\rm{3D}}(\theta_{B_{\rm loc}},\theta_{\delta z^\pm_\perp}) \propto r^{\zeta_n(\theta_{B_{\rm loc}},\theta_{\delta z^\pm_\perp})}.
\eeq
Furthermore, we define $\zeta^\parallel_n$, $\zeta^\mathrm{fluc}_n$ and $\zeta^\perp_n$ as the scaling exponents for the parallel, fluctuation-direction, and perpendicular structure functions respectively, as defined in the previous section. The scalings for the second-order structure functions were 
\beq
\zeta^\perp_2=0.50\pm0.03,\quad
\zeta^\mathrm{fluc}_2=0.69\pm0.03,\quad
\zeta^\parallel_2=0.98 \pm0.03,
\eeq
where the errors indicate standard deviations from the mean exponent obtained using the 10 snapshots. 
Thus, the turbulence exhibits significant scaling anisotropy (i.e., different scalings) with respect to all three directions identified here. The exponent in the parallel direction is very close to $1$, in good agreement with the critical-balance scaling  from Eq. (\ref{eq:eps}). Thus, the parallel scaling $\zeta^\parallel_2$ is consistent with the critical-balance conjecture. The difference between $\zeta^\perp_2$ and $\zeta^\mathrm{fluc}_2$ is consistent with the idea that the turbulent fluctuations become progressively more aligned as they cascade to smaller scales (cf. \citealt{boldyrev}, who predicts $\zeta^\perp_2=1/2$ and $\zeta^\mathrm{fluc}_2=2/3$).

\begin{figure}
\vskipfig
\includegraphics[width=8.9cm]{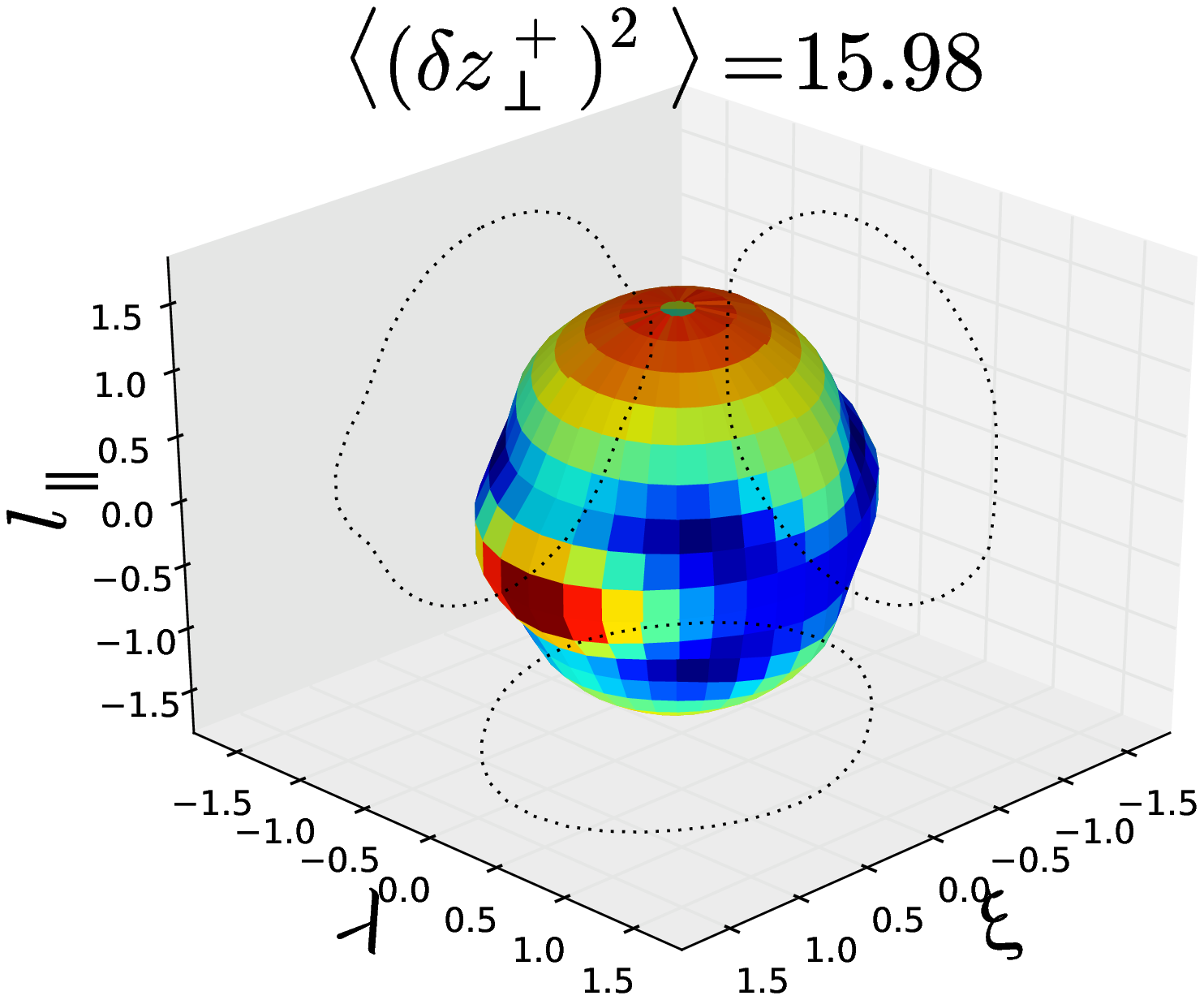}
\vskipfig
\includegraphics[width=8.9cm]{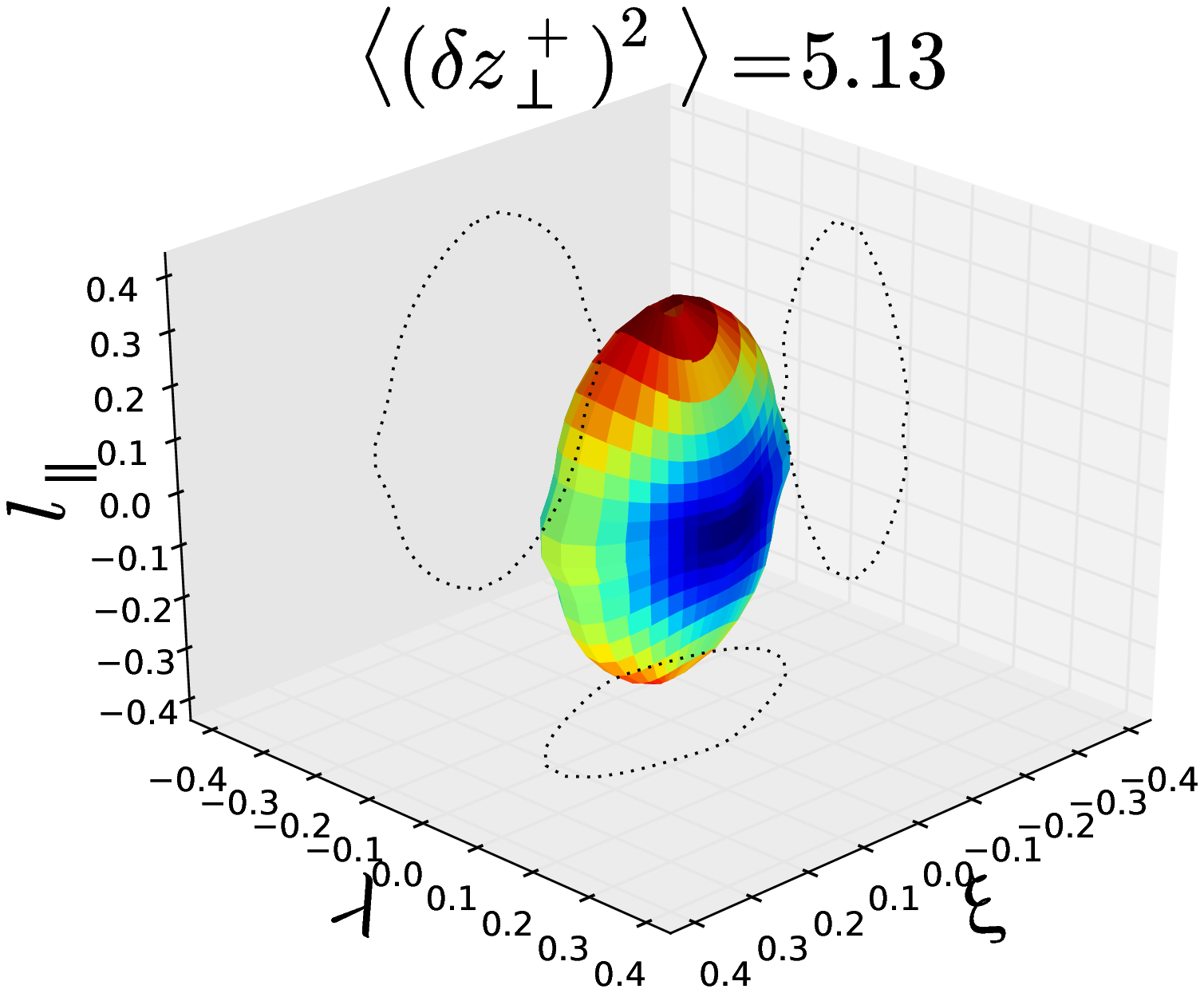}
\vskipfig
\includegraphics[width=8.9cm]{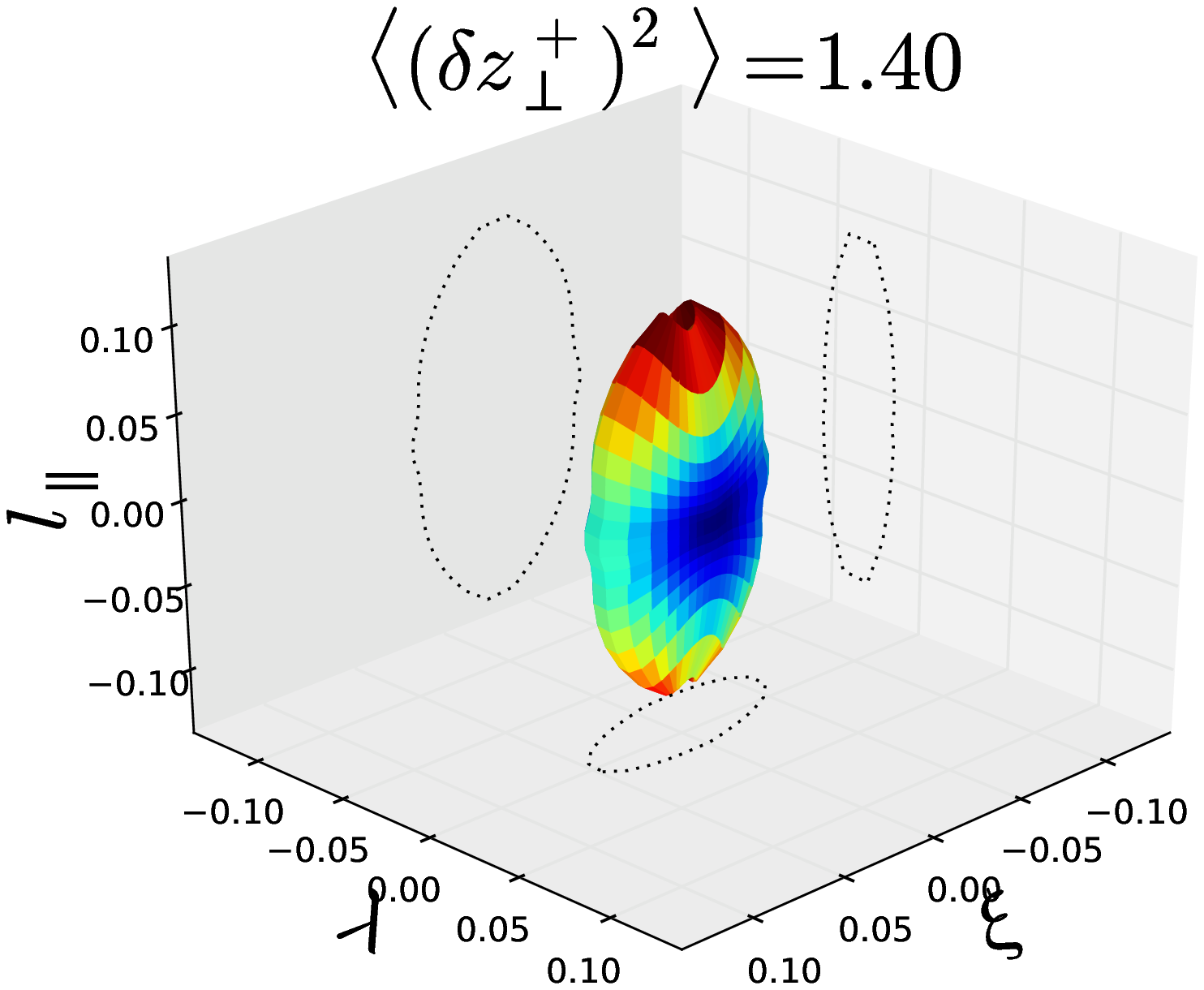}
\vskipfig
\caption{From top to bottom, ``statistical eddies" (surfaces of constant second-order structure function) at structure function values corresponding to the outer scale, roughly halfway down the inertial range, and at the bottom of the inertial range, respectively. These values are shown as three horizontal dotted lines in Figure \ref{fig:s2allang}(a). \label{fig:eddy}}
\vskipfig
\end{figure}

Based on surfaces of constant second-order structure function [Eq. (\ref{eq:s3d})], one can visualize a "statistical eddy", showing the 3D structure of turbulent correlations, in the same manner as was done for solar-wind data by \cite{chen3d}. This is done in Figure \ref{fig:eddy} for structure-function values corresponding to the outer scale, midway through the inertial range, and near the bottom of the inertial range. Statistically, due to the isotropic forcing, the structures at large scales are isotropic with respect to the local basis\footnote{Note that the parallel and perpendicular correlation lengths appear similar only because we use the nominal box aspect ratio 1. This can be arbitrarily rescaled in RMHD [see Eq.~(\ref{eq:rescaling})].}, but become increasingly "pancake"-, or "ribbon"-like deeper in the inertial range. 

One might expect some level of anisotropy imposed by constraints due to the solenoidality of the Elsasser fields. This issue is discussed in the Appendix, where we find that the solenoidality does not directly constrain the conditional structure function. 

\section{3D intermittency}\label{sec:3dinterm}
\begin{figure}
\centering
\includegraphics[width=8.9cm]{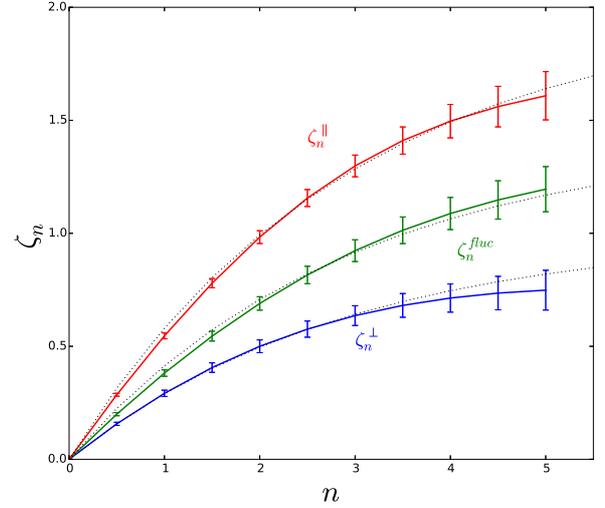}
\vskipfig
\caption{The exponents of conditional $n$th-order structure functions [Eq.~(\ref{eq:s3d})] in the perpendicular (blue), fluctuation (green) and parallel (red) directions, $\zeta_n^\perp$, $\zeta_n^\mathrm{fluc}$ and $\zeta_n^\parallel$ respectively. The lowest black dotted line is Eq. (\ref{eq:perpzeta}) with $\beta=0.71$, and the middle and highest black dotted lines are that equation with the same $\beta$, divided by $\gamma=0.69$ and $\alpha=0.49$, respectively.\label{fig:zetan}}
\vskip-0.3cm
\end{figure}

As we showed in the Introduction, to exhibit scaling anisotropy within the local perpendicular plane as seen in Section \ref{sec:2ord}, Elsasser fields must have non-self-similar scale-dependent probability distribution functions. In this section, we study the scale dependence of the distribution functions of the field increments: the intermittency of the conditional structure functions, as a function of the two local angles $\theta_{B_\mathrm{loc}}$ and $\theta_{\delta z^+}$. One common measure of intermittency is the nonlinear dependence on $n$ of the exponents of the $n$th order structure functions. We extend this approach by measuring $\zeta^\parallel_n$, $\zeta^{\rm fluc}_n$ and $\zeta^\perp_n$ --- the exponents of the parallel, fluctuation-direction, and perpendicular conditional structure functions defined in Section \ref{sec:3dsf}. These exponents are shown in Figure \ref{fig:zetan} up to $n=5$. 
An immediate conclusion is that not only is RMHD turbulence intermittent, but it is perhaps differently intermittent in all three directions. We will study these scalings in more detail in this section.

Recently, a new model of the intermittency of Alfv\'enic turbulence has been proposed by \cite{Chandran14}. The model involves two archetypal nonlinear interactions. Firstly, occasional \emph{balanced collisions} between structures of similar amplitudes $\delta z^+_\perp \sim \delta z^-_\perp$ reduce the field amplitudes. 
This motivates assuming
a log-Poisson distribution for $\delta z^\pm_\perp$. Secondly, in \emph{imbalanced collisions} with $\delta z^\pm_\perp \gg \delta z^\mp_\perp$, the amplitudes of the fluctuations remain constant while the lower-amplitude field is sheared into alignment and 
its perpendicular scale $\lambda$ reduced. The model incorporates critical balance and dynamic alignment, and predicts that the perpendicular structure function exponents are
\beq
\zeta^\perp_n = 1-\beta^n,\label{eq:perpzeta}
\eeq
where $\beta\approx0.691$ is derived via a number of assumptions. Let us fit our perpendicular exponents to this formula, and determine $\beta$ from the fit. The best-fit value\footnote{Based on minimizing the sum of the squared residuals weighted by the standard deviation of the measurements at each order \citep{aitken1936}. The errors are evaluated by varying $\beta$ until the curve given by Eq.~(\ref{eq:perpzeta}) fails to fall completely within the error bars of the measurements of the individual exponents.} is 
\beq
\beta=0.71^{+0.01}_{-0.02},
\eeq
which is in remarkably good agreement with the \cite{Chandran14} model.

It is clear from Figure \ref{fig:zetan} that the turbulence is also highly intermittent in the fluctuation (green curve) and parallel (red curve) directions. A natural question is how the parallel and fluctuation-direction scalings are related to the perpendicular ones. \cite{Chandran14} make no prediction for $\zeta_n^\parallel$ or $\zeta_n^\mathrm{fluc}$. Suppose that, in some detailed sense, the parallel and fluctuation-direction coherence scales $l_\parallel$ and $\xi$ of each turbulent fluctuation themselves have power-law dependence on the perpendicular scale $\lambda$ of that fluctuation, viz.,
\beq
l_\parallel \sim \lambda^\alpha,\quad
\xi \sim \lambda^\gamma. \label{eq:anissimple}
\eeq
In our terminology, this is equivalent to stating that the degree of anisotropy between the parallel and perpendicular or fluctuation and perpendicular directions is not itself intermittent, meaning that the scaling of the aspect ratios 
\beq
\sin\phi = \frac{\lambda}{l_\parallel} \propto \lambda^{1-\alpha}, \quad \sin\theta = \frac{\lambda}{\xi} \propto \lambda^{1-\gamma},\label{eq:aliang}
\eeq
is independent of the amplitude of the fluctuations. This is the same as conjecturing the following relationships between the scaling exponents of structure functions in different directions:
\beq
\zeta^\parallel_n = \frac{\zeta^\perp_n}{ \alpha}, \quad \zeta^{\rm fluc}_n = \frac{\zeta^\perp_n}{\gamma}.\label{eq:anisc}
\eeq
From the measured scaling exponents in Figure \ref{fig:zetan}, we find that the best-fit values are
\beq
\alpha=0.49, \quad \gamma=0.69.
\eeq
The resulting "model" curves in Eq.~(\ref{eq:anisc}) are also plotted on Figure \ref{fig:zetan}, with Eq.~(\ref{eq:perpzeta}) used for $\zeta_n^\perp$. While the curves in Eq. (\ref{eq:anisc}) are relatively close to the measured scalings, the quality of the fits is worse than in the perpendicular direction --- the model curves are not within the error bars for every $n$ measured, for any values of $\alpha$ or $\gamma$. This implies that the 
characteristic aspect ratios in Eq. (\ref{eq:aliang}) have scalings that depend on the amplitude of the fluctuations, if perhaps only slightly. In the next section, we quantify this dependence and argue that it makes physical sense.

\section{Intermittency of anisotropy}\label{sec:ass}
The dependence of the scaling of the aspect ratios 
 in Eq.~(\ref{eq:aliang}) on the amplitude of the fluctuations is a symptom of \emph{intermittency of anisotropy}: the anisotropy cannot simply be rescaled in a uniform way because fluctuations with different amplitudes at the same scale will have different typical aspect ratios. 
If we accept that perpendicular anisotropy is related to alignment as argued in the Introduction [Eq.~(\ref{eq:anisalirelation})], this is consistent with the physical model of the nonlinear interactions by \cite{Chandran14}, according to which, in an imbalanced collision, the $\vzp^+$ and $\vzp^-$ fields align to within an angle inversely proportional to the amplitude of the higher-amplitude fluctuation. This is also consistent with the finding of \cite{rcb} that the alignment angle between $\delta \vzp^+$ and $\delta \vzp^-$ is anticorrelated with the amplitude at each scale. In fact, recalling that \cite{rcb} found the critical-balance parameter
\beq
\chi^\pm \doteq \frac{\tA}{\tnl} \doteq
\frac{\lpar^\pm\dzp^\mp\sin{\theta^\pm}}{v_{\rm A}\lambda} \sim \frac{\dzp^\mp\sin{\theta}}{v_{\rm A}\sin\phi}
\label{eq:chi}
\eeq
to have a very precisely scale-invariant distribution, we realise that the simple model in Eqs. (\ref{eq:anissimple}-\ref{eq:aliang}) cannot be strictly correct: since we know that $\dzp^\mp$ is intermittent (non-scale-invariant), at least one of $\sin\theta$ and $\sin\phi$ must also be intermittent for the distribution of $\chi^\pm$ to be scale-invariant.

\begin{figure}
\vskipfig
\includegraphics[width=8.9cm]{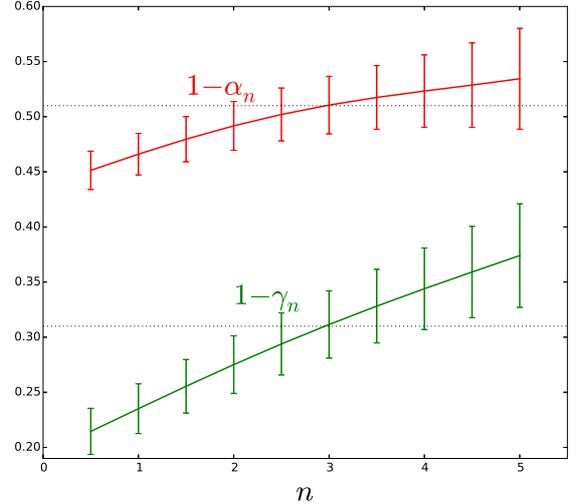}
\vskipfig
\caption{The 
aspect ratio scaling exponents $1-\alpha_n$ (red, solid line) and $1-\gamma_n$ (green, solid line) as a function of $n$ [Eq.~(\ref{eq:aliang2})]. Error bars show the standard deviation of the mean calculated from the ten snapshots. Plotted in dotted black lines are constants $1-\alpha=0.51$ and $1-\gamma=0.31$, defined in Eq. (\ref{eq:anisc}) and used for the dotted curves in Figure \ref{fig:zetan}. \label{fig:anis}}
\end{figure}

We quantify the intermittency of the anisotropy by generalising Eq.~(\ref{eq:anisc}) to
\beq
\zeta^\parallel_n = \frac{\zeta^\perp_n}{ \alpha_n}, \quad \zeta^{\rm fluc}_n = \frac{\zeta^\perp_n}{\gamma_n}.
\eeq
Then the 
aspect-ratio scalings inferred from the $n$th-order conditional structure function scalings using the equation above are given by
\beq
\sin\phi_n \propto \lambda^{1-\alpha_n}, \quad \sin\theta_n \propto \lambda^{1-\gamma_n},\label{eq:aliang2}
\eeq
so we are now allowing some amplitude dependence of these scalings. Figure \ref{fig:anis} shows these scalings as a function of $n$. Both $\sin\phi_n$ and $\sin\theta_n$ have scaling exponents that increase with $n$, meaning that the fluctuation amplitude and 
$\sin\theta$ (and, therefore, fluctuation amplitude and the alignment angle as measured by, for example, $\sin\theta^\pm$) are anticorrelated at each scale, confirming the result of \cite{rcb} and the expectation based on the physical picture of nonlinear interactions in the model of \cite{Chandran14}. 
From the range of variation exhibited by $\alpha_n$ and $\gamma_n$ in Figure \ref{fig:anis}, we conclude that the parallel aspect ratio $\sin\phi_n$ exhibits only slight intermittency, while the perpendicular aspect ratio $\sin\theta_n$ is more significantly intermittent. Note, however, that the slight variation of $1-\alpha_n$ with $n$ is nevertheless likely to be real: \cite{rcb} found that the nonlinear time alone [Eq.~(\ref{eq:tnl})] was not as precisely scale-invariant as $\chi^\pm$ [Eq. (\ref{eq:chi})].
\begin{figure*}
\begin{tabular}{cc}
\includegraphics[width=8.9cm]{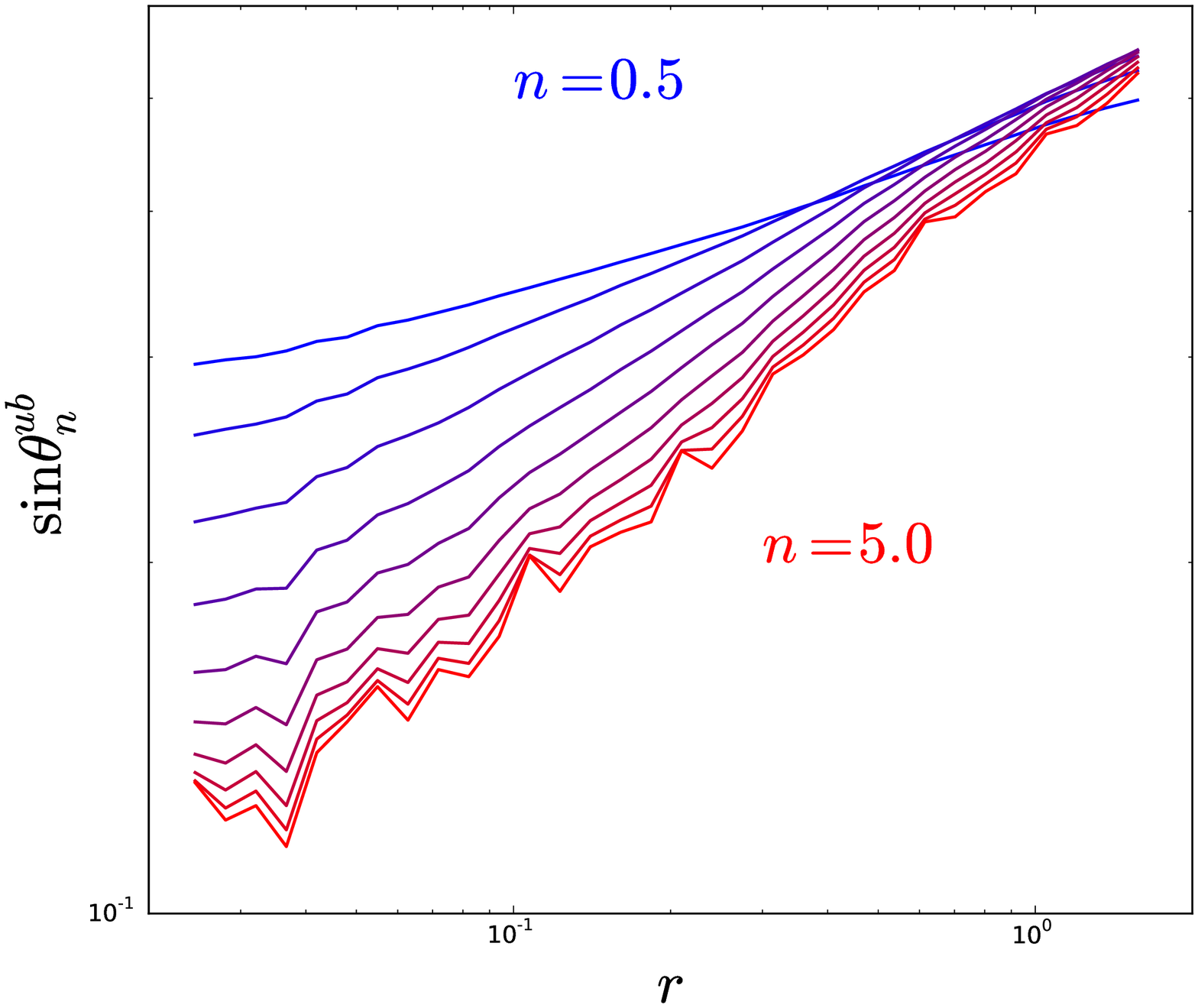} &
\includegraphics[width=8.9cm]{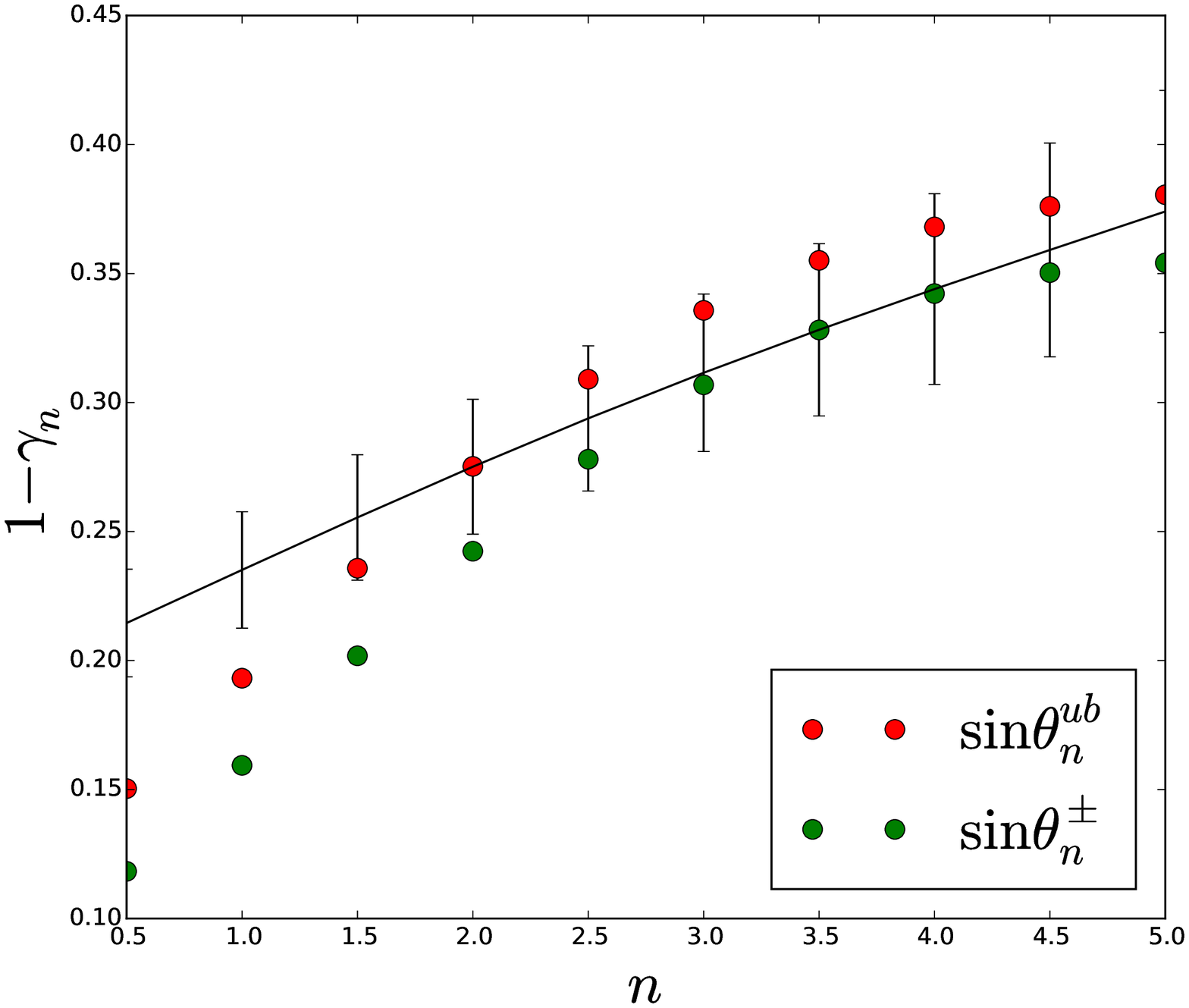} \\
\textbf{(a)} & \textbf{(b)} \\
\end{tabular}
\caption{\textbf{(a)} The alignment measures $\sin{\theta}^{ub}_{n}$ [Eq.~(\ref{eq:thubln})] are plotted, going from $n=0.5$ (blue) to $n=5$ (red). \textbf{(b)} Comparison between the scaling exponents of $\sin{\theta}^{ub}_{n}$ (red points), the scaling exponents of $\sin{\theta}^{\pm}_{n}$ (green points) and the perpendicular alignment exponents $(1-\gamma_n)$ (black line) defined in Eq. (\ref{eq:aliang2}) (and plotted in Figure \ref{fig:anis}).\label{fig:thsf}}
\end{figure*}
\section{Comparison between different measures of alignment}\label{sec:comp}
Had $\gamma_n$ been independent of $n$ (i.e., had 
the perpendicular aspect ratio $\sin\theta_n$ been non-intermittent), the alignment within the sheetlike structures would also have been non-intermittent, and it would not have mattered what method one used to measure the scaling of the alignment angle. But $\gamma_n$ is intermittent, and so the precise measure of alignment does matter. \cite{mcatbolalign} calculated $\theta^{ub}_n$ defined by
\beq
\sin^n{\theta}^{ub}_{n} \equiv \frac{\langle|\delta \vu_{\perp} \times \delta \vb_{\perp}|^n\rangle}{\langle|\delta \vu_{\perp}|^n|\delta\vb_{\perp}|^n\rangle}, \label{eq:thubln}
\eeq
with $n=1$, and found that $\theta^{ub}_1 \sim \lambda^{0.25}$, which they interpreted as vindication of the \cite{boldyrev} phenomenological theory (which was not concerned with intermittency). In contrast, \cite{blpolinterm} measured 
\beq
\sin\tilde{\theta} = \left\langle \frac{|\mathbf{\delta z^+_\perp} \times \mathbf{\delta z^-_\perp}|}{|\mathbf{\delta z^+_\perp} || \mathbf{\delta z^-_\perp}|}\right\rangle,
\eeq
and found that this quantity exhibited virtually no scale dependence, showing that how one weights the angle by the amplitude of the fluctuation matters a great deal. We may also define another set of measures of alignment, via 
\beq
\sin^n{\theta}^{\pm}_{n} \equiv \frac{\langle|\delta \vz^+_{\perp} \times \delta \vz^-_{\perp}|^n\rangle}{\langle|\delta \vz^+_{\perp}|^n|\delta\vz^-_{\perp}|^n\rangle}. \label{eq:thpmln}
\eeq
Figure \ref{fig:thsf} shows the scale dependence of $\sin\theta^{ub}_n$ for $0.5 \leq n \leq 5$. The fact that the scaling of these alignment measures depends on how one weights them with amplitude is consistent with the idea that the alignment angle and amplitude are anticorrelated at each scale. In the foregoing, we calculated $\sin\theta_n \propto \lambda^{(1-\gamma_n)}$ in terms of the scalings of the 3D conditional structure function, and argued that $\theta_n \sim \theta^{ub}_n \sim \theta_n^\pm$. In Figure \ref{fig:thsf}, the scaling exponents of $\sin\theta^{ub}_n$ and and $\sin\theta_n^\pm$ are compared with $1-\gamma_n$, where $\gamma_n$ are the perpendicular alignment exponents 
defined in Eq. (\ref{eq:aliang2}) and plotted in Figure \ref{fig:anis}. The agreement is not perfect, but the three different measures show the same trend, and agree at high $n$, suggesting, as we argued in the Introduction, that the same physical phenomenon is being measured using our technique as in previous work.

\section{Discussion}\label{sec:disc}
The results presented in this paper show that strong Alfv\'enic turbulence scales highly anisotropically with respect to all three physically relevant directions: parallel ($\hat{\vB}_{\rm loc}$), fluctuation ($\delta \hat{\vz}_\perp^\pm$), and perpendicular ($\delta \hat{\vz}_\perp^\pm \times \hat{\vB}_{\rm loc}$). This anisotropy can be explained using two key physical ideas. The critical-balance conjecture underpins the parallel anisotropy, while the anisotropy within the perpendicular plane 
can be linked with scale-dependent alignment of the fluctuations.

The intermittent scalings $\zeta_n^\perp, \zeta_n^\mathrm{fluc},\zeta_n^\parallel$ of the conditional structure functions in these three directions reported here shed further light on the physics of critical balance and alignment. The perpendicular scalings agree closely with the predictions of the model of \cite{Chandran14}. The 
aspect-ratio scaling $\sin\theta_n = \lambda/\xi \propto \lambda^{1-\gamma_n}$ can be inferred from the ratio of the scaling exponents of the perpendicular- and fluctuation-direction structure functions, $\gamma_n = \zeta_n^\perp/\zeta_n^\mathrm{fluc}$, and we find that the scaling exponent $1-\gamma_n$ is an increasing function of $n$. This implies that the alignment angle is anticorrelated with amplitude at each scale, i.e., the alignment angle is intermittent (not scale-invariant). This promotes the view that alignment is set by mutual shearing of the Elsasser fields, which naturally leads to such anticorrelation \citep{Chandran14}. Meanwhile, the scaling of the aspect ratio 
between the perpendicular and parallel directions, $\sin\phi_n \propto \lambda^{1-\alpha_n}$ varies only slightly with $n$ (although the results of \cite{rcb} suggest that this is variation is real).%

In the solar wind, \cite{chen3d} applied the 3D conditional structure function technique and found essentially scale-independent anisotropy between the perpendicular and fluctuation directions in fast solar wind. 
\cite{wicks} also found essentially no scaling of the alignment angle in the inertial range of the fast solar wind.  \cite{Chandran14} provide a review of various different solar-wind measurements, showing that there appears to be a significant spread in the measured structure-function exponents, possibly depending on whether the measurement was from the fast or slow solar wind. This could also affect the measurement of the alignment. The difference between the fast-solar-wind measurements in \cite{wicks, chen3d} and our simulations appears to be the presence of significant  anisotropy within the perpendicular plane (or equivalently, alignment) at the outer scale in the solar wind, but not in our simulations. This difference is also evident in \cite{verdini2015}, who link differences in the anisotropy of conditional structure functions to the expansion of the solar wind. The effect of this expansion can clearly be seen at the outer scale of their expanding box simulation, i.e. at scales where the dynamics of the cascade have not yet affected the anisotropy, but at the smaller scales anisotropy similar to that measured here is observed. Finally, \cite{Osman14} have previously considered the intermittency of the parallel structure functions in the solar wind using the conditional structure function method, but obtained different results to those presented here. The reason for this difference requires further investigation. 

Further measurements of the anisotropy and intermittency in the solar wind and of the dependence of the intermittent scalings in all directions on the solar-wind conditions would allow for new comparisons between the real turbulence and the numerical simulations presented here, and improve our understanding of the physical processes underlying dynamic alignment, critical balance, and intermittency. What appears to be suggested by the detailed study undertaken here is that all of these phenomena are very much intertwined.

\section*{Acknowledgements}
This work was supported in part by NASA grant NNX15AI80G and NSF grant PHY-1500041. Simulations reported here used XSEDE, which is supported by the US NSF Grant ACI-1053575. We acknowledge support provided to ISSI/ISSI-BJ Team 304. C. H. K. Chen is supported by an Imperial College Junior Research Fellowship. 

\appendix
\section{Solenoidality and perpendicular anisotropy}
Since the Elsasser fields are 2D solenoidal, $\nabla_\perp \cdot \vzp^\pm =0$, one might expect some degree of anisotropy with respect to the fluctuation direction within the perpendicular plane due to just this kinematic property. In this Appendix, we will outline the constraints that solenoidality places on the turbulence and to what extent this is related to anisotropic scalings of the conditional structure functions, Eq. (\ref{eq:s3d}).

We work within the perpendicular plane. We will use a basis for each separation $\vrpp$ where the $x$ direction points along $\vrpp$ and the $y$ direction is transverse to it. Since $\vzp^\pm$ are globally isotropic (within the $(x,y)$ plane perpendicular to the global mean magnetic field), the $n$th-order two-point structure function of $\delta \vzp^\pm$, the rank-$n$ tensor $\langle \delta z^\pm_{\perp,i} \delta z^\pm_{\perp,j} \delta z^\pm_{\perp,k}\ldots\rangle$, can be expressed as a sum of terms, each with $n$ vector indices and composed of products of Kronecker deltas $\delta_{ij}$ and unit vectors $\hat{r}_{\perp,i}$, all with distinct indices. Moreover, the structure function must be of such a form that it is invariant under interchange of indices. For example, the tensor second-order structure function is
\beq
\langle \delta z^\pm_{\perp,i} \delta z^\pm_{\perp,j}\rangle = S_{2,\rm T}(r_\perp)(\delta_{ij} - \hat{r}_{\perp,i}\hat{r}_{\perp,j}) + S_{2,\rm L}(r_\perp)\hat{r}_{\perp,i}\hat{r}_{\perp,j},\label{eq:full2ord}
\eeq
where $r_\perp = |\vrpp|$ and the longitudinal $S_{2,\rm L}$ and transverse $S_{2,\rm T}$ scalar structure functions are
\beq
\begin{split}
S_{2,\rm L} = \langle (\delta \vzp^\pm \cdot \hat{\mathbf{r}}_\perp)^2 \rangle, \quad S_{2,\rm T} = \langle (\delta \vzp^\pm\! \times \hat{\mathbf{r}}_\perp)^2 \rangle. 
\end{split}
\eeq
The solenoidality constraint is imposed by taking the divergence $\partial/\partial r_i$ of Eq. (\ref{eq:full2ord}) and setting it equal to zero. This gives the von K\'arm\'an relation in 2D \citep{batchelor1953}:
\beq
 \frac{\partial}{\partial r_\perp} \left(r_\perp S_{2,\rm L}\right) = {S_{2,\rm T}}.\label{eq:solconst}
\eeq
This means that in the inertial range, where $S_{2,\rm L}$ and $S_{2,\rm T}$ are power laws, they must scale in the same way $\propto r_\perp^{2a}$, and have a certain ratio $(2a+1)$ between them:
\beq
S_{2,\rm T} = D r_\perp^{2a}, \quad S_{2,\rm L} = \frac{D}{2a+1} r_\perp^{2a},\label{eq:sll}
\eeq
where $D$ is a constant. Thus there is a scale-independent level of "kinematic" anisotropy between the transverse and longitudinal structure functions. 

The third order tensor structure function again depends on two scalar functions of $r_\perp$, each multiplying one of the only two possible rank-$3$ tensors, $\hat{r}_{\perp,i}\hat{r}_{\perp,j}\hat{r}_{\perp,k}$ and $\delta_{ij}\hat{r}_{\perp,k}+\delta_{jk}\hat{r}_{\perp,i}+\delta_{ki}\hat{r}_{\perp,j}$. Solenoidality again amounts to setting the divergence equal to zero, and gives a homogenous constraint that guarantees that all components of the third-order structure function are either zero or have the same scaling, but allows for scale-independent ratios between the components, similar to the second-order case.
At higher orders than 3, there are no more solenoidality constraints, because structure functions contain terms such as $\langle z^\pm_{\perp,i}(0)z^\pm_{\perp,j}(0)z^\pm_{\perp,k}(\mathbf{r}_\perp)z^\pm_{\perp,l}(\mathbf{r}_\perp)\rangle$ whose divergence does not vanish \citep{lvov1997,hill2001}.

Using the second-order structure function as an example, the longitudinal and transverse structure functions $S_{2,\rm{L}}$, $S_{2,\rm{T}}$ are not directly related to the conditional structure function $S_{2,3D}$ defined in Eq. (\ref{eq:s3d}). $S_{2,\rm{L}}$ and $S_{2,\rm{T}}$ are moments of the joint distribution $p(\delta z^\pm_\perp,\theta_{\delta z^\pm_\perp} | r_\perp)$ of the field-increment amplitude $\delta z^\pm_\perp$ \emph{and} the angle $\theta_{\delta z^\pm_\perp}$, conditional on the separation distance $r_\perp$, viz.,
\begin{equation}
\begin{split}
S_{2,\rm{L}}(r_\perp) = \int_0^\infty \int_0^{2\pi} (\delta z^\pm_\perp)^2 \cos^2(\theta_{\delta z^\pm_\perp}) \\ \times p(\delta z^\pm_\perp,\theta_{\delta z^\pm_\perp} | r_\perp) d\delta z^\pm_\perp d\theta_{\delta z^\pm_\perp}, \\
S_{2,\rm{T}}(r_\perp) = \int_0^\infty \int_0^{2\pi} (\delta z^\pm_\perp)^2 \sin^2(\theta_{\delta z^\pm_\perp}) \\\times p(\delta z^\pm_\perp,\theta_{\delta z^\pm_\perp} | r_\perp) d\delta z^\pm_\perp d\theta_{\delta z^\pm_\perp}.
\end{split}
\end{equation}
We have shown that these functions must have the same scaling due to solenoidality. In contrast, the conditional structure function defined by Eq. (\ref{eq:s3d}) (ignoring for now the dependence on the third dimension via the angle $\theta_{B_{\rm loc}}$) is the moment of the distribution of the field increment amplitude $\delta z^\pm_\perp$ \emph{conditional on} the angle $\theta_{\delta z^\pm_\perp}$ and the separation distance $r_\perp$:
\beq
S_{2,\rm{3D}}(r_\perp,\theta_{\delta z^\pm_\perp}) = \int_0^\infty (\delta z^\pm_\perp)^2 p(\delta z^\pm_\perp | \theta_{\delta z^\pm_\perp},r_\perp) d\delta z^\pm_\perp.
\eeq
Thus, in general, $S_{2,\rm 3D}$ coincides with neither $S_{2,\rm{L}}$ at $\theta_{\delta z^\pm_\perp}=0$ nor with $S_{2,\rm{T}}$ at $\theta_{\delta z^\pm_\perp}=\pi/2$. Therefore, the scale-dependent anisotropy of the turbulence within the perpendicular plane as measured by $S_{2,\rm 3D}$ in Figure \ref{fig:s2allang} cannot be expressed simply in terms of the solenoidality constraints. 



\bibliographystyle{mnras}
\bibliography{mainbib} 


\bsp	
\label{lastpage}
\end{document}